# Large-Eddy Simulations of Stochastic Burgers Turbulence Using Fully Conservative Higher-Order Schemes

Mehran Sharifi*

*Department of Mechanical Engineering, Amirkabir University of Technology, Iran*

**Corresponding Author**
Mehran Sharifi, Department of Mechanical Engineering, Amirkabir University of Technology, Iran.

**Abstract**

*The study of Large-Eddy Simulations (LES) in turbulent flows continues to be a critical area of research, particularly in understanding the behavior of small-scale turbulence structures and their impact on resolved scales. In this study, we focus on the LES of turbulent flows, particularly the one-dimensional Stochastic Burgers Equation (SBE), using fully conservative higher-order schemes. The interaction between spatial discretization and SubGrid-Scale (SGS) modeling is explored rigorously by validating these schemes against analytical solutions for both the Linear Advection-Diffusion (LAD) equation and the Non-Linear Burgers (NLB) equation under laminar conditions. This ensures robustness before applying the approach to LES of stochastic turbulence. The study investigates how second-order and fourth-order discretization schemes influence the dynamic coefficients of various SGS models, including Constant Smagorinsky (CS), Dynamic Smagorinsky (DS), Dynamic Wong-Lilly (DWL), 1.5-order Turbulent Kinetic Energy Deardorff (TKED), Equilibrium Heinz (EH), and Dynamic Heinz (DH) models. The second-order scheme was found to amplify fluctuations in dynamic SGS coefficients due to its higher numerical dissipation, contrasting with the more stable behavior observed with the fourth-order scheme, which better captures resolved scales and results in smaller dynamic coefficients. Despite inherent differences in SGS models and discretization schemes, the final velocity distributions in one-dimensional turbulence simulations were remarkably consistent, suggesting a limited influence of SGS modeling on large-scale structures in simplified turbulence scenarios. However, notable variations in resolved-scale kinetic energy were uncovered, emphasizing the importance of accurately capturing small-scale turbulence structures for precise energy dissipation predictions in LES.*

## Nomenclature

**Latin symbols**

| | | |
|---|---|---|
| $C_{CS}$ | : | Coefficient of Constant Smagorinsky, – |
| $C_{DH}$ | : | Coefficient of Dynamic Heinz, – |
| $C_{DS}$ | : | Coefficient of Dynamic Smagorinsky, – |
| $C_{DWL}$ | : | Coefficient of Dynamic Wong-Lilly, – |
| $C_{EH}$ | : | Coefficient of Equilibrium Heinz, – |
| $D_0$ | : | Noise amplitude, $m^2\ s^{-3}$ |
| $f(x)$ | : | Gaussian random variable, – |
| $\hat{f}(k_w)$ | : | Fourier transform, – |
| $F^{-1}$ | : | Inverse Fourier transform, – |
| $G(r)$ | : | One-dimensional box filter, – |
| $H(\chi)$ | : | Heaviside function, – |
| $k$ | : | SGS turbulent kinetic energy, $J$ |
| $k_w$ | : | Wavenumber, $m^{-1}$ |
| $KE$ | : | Resolved-scale kinetic energy, $J$ |
| $K_{DH}$ | : | Total coefficient Dynamic Heinz, $J^{\frac{1}{2}} m$ |
| $K_{TKED}$ | : | Coefficient of Turbulent Kinetic Enertgy Deardorff, $J^{\frac{1}{2}} m$ |
| $l$ | : | Domain length, $m$ |
| $L_{11}$ | : | Resolved stress (or Germano Identity), $m^2\ s^{-2}$ |
| $M_{11}$ | : | Scaled composite rate-of-strain, $m^2\ s^{-2}$ |
| $M_{11}^*$ | : | Modified scaled composite rate-of-strain, $m^2\ s^{-2}$ |
| $N$ | : | Number of grid points, – |
| $N_{11}$ | : | Scaled gradient of velocity, $m^{4/3}\ s^{-1}$ |
| $t$ | : | Time, $s$ |
| $u(x,t)$ | : | One-dimensional velocity field, $m\ s^{-1}$ |

**Greek symbols**

| | | |
|---|---|---|
| $\beta$ | : | Spectral slope of the noise, – |
| $\Delta_f$ | : | Filter width, $m$ |
| $\Delta_t$ | : | Test width, $m$ |
| $\eta(x,t)$ | : | Noise term, $m\ s^{-2}$ |
| $\nu$ | : | Kinematic viscosity, $m^2\ s$ |
| $\tau^R$ | : | Residual (or SGS) stress, $m^2\ s^{-2}$ |
| $\phi$ | : | General quantity, – |

**Superscripts**

| | | |
|---|---|---|
| $\overline{\phantom{x}}$ | : | Filtered value |
| $\tilde{\phantom{x}}$ | : | Test filtered value |
| CS | : | Constant Smagorinsky Model |
| DH | : | Dynamic Heinz Model |
| DS | : | Dynamic Smagorinsky Model |
| DWL | : | Dynamic Wong-Lilly Model |
| EH | : | Equilibrium Heinz Model |
| TKED | : | Turbulent Kinetic Energy Deardorff Model |

## 1. Introduction

Large-eddy simulation (LES) is a sophisticated Computational Fluid Dynamics (CFD) technique that has garnered significant attention in the study of turbulent flows across various engineering and scientific applications. By resolving large-scale turbulent structures directly and modeling the effects of smaller scales through SubGrid-Scale (SGS) models, LES offers a more detailed representation of turbulent flows compared to traditional Reynolds-Averaged Navier–Stokes (RANS) models [1,2]. This methodology enables researchers to capture the most significant scales of turbulence, providing valuable insights into complex flow phenomena, as extensively elaborated by Okraschevski [3]. The development of LES methodologies has been an active area of research, focusing on enhancing both the accuracy and efficiency of turbulent flow simulations. Caban and Tyliszczak explored advanced filtering techniques to improve the modeling of turbulent reactive flows in combustion applications [1]. Additionally, Tian et al. investigated the incorporation of physics-informed machine learning approaches to enhance the predictive capabilities of Lagrangian LES simulations [4]. Such advancements highlight the continuous evolution of LES techniques in addressing challenges associated with turbulent flow modeling. One of the key advantages of LES is its ability to capture unsteady features of flows and aero-acoustic properties, as demonstrated by Lombard et al. in their study of wingtip vortices [5]. The application of LES extends into various fields, including aerospace engineering and environmental fluid dynamics. For instance, Li et al. demonstrated the use of LES in studying hurricane evolution during landfall, enabling an investigation into the impact of complex flow structures, such as roll vortices, on atmospheric behavior [6]. Furthermore, LES has been applied to analyze wind turbine wakes and their sensitivity to SGS models, showcasing its versatility in studying aerodynamic phenomena [7]. In combustion research, Ali et al. emphasized the pivotal role of LES in studying turbulent premixed combustion processes under varying conditions [8]. By simulating reactive flows in combustion chambers, LES provides detailed insights into the interactions between turbulence and combustion dynamics, leading to advancements in combustion system design and optimization. This intricate interplay between flow dynamics and chemical reactions has solidified LES's position as a valuable tool in propulsion research and environmental sustainability efforts. The integration of LES with other computational methods has further expanded its applicability in diverse fields. Zhang and Shih developed hybrid LES-RANS approaches, effectively combining the strengths of both methodologies for more efficient and accurate flow simulations [9]. Moreover, Kempf and Munz demonstrated the use of zonal direct-hybrid simulations to enable simultaneous modeling of large eddy dynamics and acoustic propagation, illustrating the versatility of LES in aeroacoustics studies [10]. The advancements in LES methodologies are also evident in industrial applications aimed at optimizing processes involving fluid dynamics and heat transfer. Li et al. utilized LES to predict conjugate heat transfer in turbulent channel flows, demonstrating its accuracy in capturing thermal processes [2]. Additionally, Korinek and Tisovský employed LES in modeling impinging jet heat transfer, evaluating the influence of local grid refinement on heat transfer characteristics, showcasing its utility in various industrial heat transfer applications [11]. Furthermore, Hu et al. utilized LES to explore the formation of vortices around hydrofoils, while Protas et al. highlighted the limitations of RANS models, where turbulence is modeled empirically, in accurately capturing turbulent structures [12,13]. By resolving larger turbulent structures, LES provides a more detailed representation of the flow field, making it invaluable for understanding turbulence dynamics [14]. The evolution of LES has been marked by advancements in numerical methods and modeling techniques, with Östh et al. exploring nonlinear subscale turbulence terms to enhance model accuracy, particularly in high-Reynolds-number flows [15]. Xie et al. further developed reduced-order modeling techniques that leverage LES data for more efficient simulations of turbulent flows, underscoring the continuous refinement of LES methodologies to improve their predictive capabilities and computational efficiency [16].

SGS models play a crucial role in LES by effectively capturing the impact of unresolved turbulent motions on the resolved scales. These models are specifically designed to represent the SGS turbulence that cannot be explicitly resolved by the computational grid. A notable characteristic of SGS models is their ability to account for the dissipation of energy from resolved to unresolved scales, as highlighted by Moser et al. [17]. Recent advancements have marked a significant shift towards data-driven approaches. For instance, Xie et al. explored the use of Artificial Neural Networks (ANNs) to develop nonlinear algebraic models for SGS stresses in LES, showcasing how these models can enhance both accuracy and efficiency by capturing complex turbulent behaviors [18]. Supporting this notion, Subel et al. demonstrated the potential of such data-driven models in refining LES results [19]. In addition, the Dynamic Smagorinsky (DS) model, introduced by Mallik et al., represents a step forward in dynamic SGS modeling [20]. This model improves predictive capabilities by adjusting coefficients based on evolving flow dynamics.

Complementing these efforts, researchers have increasingly applied machine learning techniques, such as deep learning and convolutional neural networks, to develop SGS models tailored for specific scenarios, including stable a posteriori LES of two-dimensional turbulence and predictions of wind turbine wakes, as explored by Guan and Ghobrial [7,21]. Moreover, Inagaki and Kobayashi emphasized the importance of accurately modeling the transport of SGS turbulent kinetic energy across various flow configurations [22]. Their research underlines the critical role of SGS energy flux representation in achieving realistic simulations. Additionally, Qi et al. and Yuan et al. have concentrated on developing scale-similarity dynamic procedures and quasi-dynamic SGS kinetic energy models, respectively, to enhance LES fidelity in capturing compressible flows [23,24]. As for practical applications, Vela-Martín and Long et al. have successfully customized SGS models for simulating a wide range of phenomena, including isotropic turbulence and bubble column bubbly flow [25,26]. Duben et al. also contributed to the field by investigating jet aerodynamics, demonstrating the versatility and adaptability of these models across diverse domains [27]. Also, Hickling presented promising results with the integration of deep learning techniques, such as adjoint-trained models, which indicate a trend toward more autonomous and data-informed SGS modeling approaches [28]. This evolution signifies a progressive shift in the landscape of LES, emphasizing the importance of innovative methodologies for capturing complex turbulent phenomena.

One-dimensional Stochastic Burgers Turbulence (SBT) serves as a pivotal model for understanding the intricate dynamics of turbulent flows influenced by stochastic processes. The Stochastic Burgers Equation (SBE), a type of stochastic Partial Differential Equation (PDE), captures the interplay between advection and diffusion, offering valuable insights into the behavior of turbulent systems under random perturbations. For instance, Dong et al. extensively explored the global well-posedness, regularity, and stability aspects of this model, elucidating the dynamics of turbulent flows subjected to multiplicative noise [29]. In their research, Dotsenko and De et al. investigated various characteristics of one-dimensional SBT, focusing on velocity distribution functions, intermittency, and dynamic multi-scaling [30,31]. Their findings significantly enhance our understanding of the statistical properties and scaling behaviors inherent in the turbulent flow field. Additionally, Lu advanced the field by employing data-driven model reduction techniques to develop efficient parametric closure models specifically tailored for one-dimensional SBE, thereby improving the computational efficiency and accuracy of simulations [32]. The study of ergodicity within the Burgers system has also garnered attention. Peszat et al. examined the existence of invariant measures and the long-term behavior of solutions driven by stochastic processes, further contributing to our understanding of the system's dynamics [33]. Moreover, the development of weak Galerkin Finite Element Methods (FEMs), along with the formulation of exact solutions through specialized approximation techniques, has been instrumental in the numerical treatment and analysis of coupled viscous Burgers equations. Notably, Chen and Zhang as well as Nazir et al. provided valuable insights that enhance our grasp of turbulent phenomena through their methodological advancements [34,35]. Furthermore, the exploration of moderate deviations, strong convergence properties, and large deviation principles for SBEs has shed light on rare events and long-term statistical behaviors in turbulent flows perturbed by stochastic noise. For example, Belfadli et al. Jentzen et al. and Gao all contributed to this area, deepening our understanding of the implications of stochastic perturbations on turbulence [36-38]. In addition, the connection between SBEs and optimal control theory has been thoughtfully examined by Mohan et al. [39]. Their work elucidates optimal strategies for influencing the evolution of turbulent systems under uncertainty, highlighting the practical implications of these theoretical frameworks.

In this investigation, our aim is to utilize accurate numerical methods, specifically second-order and fourth-order spatial discretization schemes, for LES of the one-dimensional SBE within the context of turbulent flow. These methods, referred to as higher-order schemes, will be implemented using Python.

The structure of the article is as follows: Sections 2.1 and 2.2 introduce the problem statement and the modeling approach. Section 2.3 details the numerical solution procedure, including the discretization of the governing equations. In Section 3, the validation process is presented, where results obtained from higher-order schemes are compared with analytical solutions. Section 4 delves into the analysis, with Section 4.1 examining the impact of discretization methods on the evolution of dynamic coefficients across different SGS models. Section 4.2 evaluates the velocity distribution at the final time, comparing results from various SGS models using both second-order and fourth-order discretization schemes to assess accuracy. Section 4.3 investigates the effects of different dynamic SGS models on the evolution of resolved-scale kinetic energy. The article concludes with a summary of the key findings in Section 5.

## 2. Physical and Mathematical Modelling
### 2.1 Problem Statement
The one-dimensional SBT model plays a crucial role in elucidating the complex dynamics of turbulent flows affected by stochastic processes. According to Basu the one-dimensional SBE, which is a form of stochastic PDE, encapsulates the interaction between advection and diffusion [40]. This provides significant insights into the behavior of turbulent systems when subject to random disturbances. The equation can be written as [40]

$$\frac{\partial u}{\partial t} + u\frac{\partial u}{\partial x} = v\frac{\partial^2 u}{\partial x^2} + \eta(x,t) \quad (1)$$

Where $u(x,t)$, $\eta(x,t)$, $v$ are the velocity field, noise term, and kinematic viscosity, respectively. This equation is defined over a domain of length $l = 2\pi$, with periodic boundary condition ($u(0,t) = u(2\pi,t)$) and an initial condition of zero ($u(x,0)=0$). The noise term, which is temporally white but spatially correlated, can be defined as follows [41]:

$$\eta(x,t) = \sqrt{\frac{2D_0}{\Delta t}} F^{-1}\left\{|k_w|^{\frac{\beta}{2}} \hat{f}(k_w)\right\} \quad (2)$$

In this context, kinematic viscosity ($v$), noise amplitude ($D_0$), the spectral slope of the noise ($\beta$), and the time step size ($\Delta t$) are set to values of $10^{-5} m^2/s$, $10^{-6} m^2/s^3$, $-\frac{3}{4}$ and $10^{-4}s$, respectively [40,42]. Furthermore, $k_w$ denotes the wavenumber, and $f(k_w)$ represents the Fourier transform of a Gaussian random variable (field) $f(x)$, which has a mean of zero and a standard deviation of $\sqrt{N}$. The notation $F^{-1}$ indicates the inverse Fourier transform, while $N$ signifies the number of grid points. According to Basu the Direct Numerical Simulation (DNS) is conducted for a duration of $t = 200s$ with a resolution of $N = 8192$, while the LES is performed for the same total time with a resolution of $N = 512$ [40]. It is important to note that for DNS, it suffices to discretize equation (1) and solve the resulting algebraic equation numerically. In contrast, LES simulations require the formulation of a filtered equation derived from equation (1), along with the implementation of SGS models. The subsequent section presents comprehensive formulations of the LES and SGS models.

### 2.2 Formulations of LES and SGS Models
The filtered version of SBE can be achieved by utilizing a one-dimensional box filter in the following manner proposed by Basu [40]:

$$\frac{\partial \bar{u}}{\partial t} + \frac{1}{2}\frac{\overline{\partial uu}}{\partial x} = v\frac{\overline{\partial^2 u}}{\partial x^2} + \bar{\eta}(x,t) \quad (3)$$

Given the assumption that the box filter is homogeneous, the derivative-filtering error approaches zero [43]. Consequently, in equation (3), the positions of the spatial derivatives and the one-dimensional box filter can be interchanged (for example: $\frac{\overline{\partial uu}}{\partial x} \sim \frac{\partial \overline{uu}}{\partial x}$ and $\frac{\overline{\partial^2 u}}{\partial x^2} \sim \frac{\partial^2 \bar{u}}{\partial x^2}$). Therefore, the final form of the filtered SBE can be expressed as follows [44]:

$$\frac{\partial \bar{u}}{\partial t} + \bar{u}\frac{\partial \bar{u}}{\partial x} = v\frac{\partial^2 \bar{u}}{\partial x^2} - \frac{1}{2}\frac{\partial \tau^R}{\partial x} + \bar{\eta}(x,t) \quad (4)$$

In equations (3) and (4), the bar denotes the filtering operation using a filter with a characteristic width of $\Delta_f = \Delta x$, as described by Geurts [45]. The filtered equation is now suitable for numerical solution through LES on a grid with a mesh size of $\Delta x$, which is significantly larger than the smallest scale of motion, known as the Kolmogorov scale. Additionally, in equation (4), $\tau^R = \overline{(uu)} - \overline{uu}$ and $\bar{\eta}(x,t)$ represent the residual (or SGS) stress and the filtered forcing function, respectively. $\tau^R$ can be obtained using various SGS models, while the filtered forcing function $\bar{\eta}(x,t)$ is defined as follows [44]:

$$\bar{\eta}(x,t) = \int_D G(r,x)\eta(x-r,t)dr \xrightarrow{\substack{Homogeneous \\ Assumption}} \bar{\eta}(x,t)$$

$$= \int_D G(r)\eta(x-r,t)dr \quad (5)$$

Given that $D$ is the computational domain and $G(r,x) \sim G(r)$ represents a one-dimensional box filter. This filter function (or kernel) can be expressed using the Heaviside function, $H(\chi)$, as follows [44]:

$$G(r) = \frac{1}{\Delta_f}H(\chi)\Big|_{\chi = \frac{\Delta_f}{2} - |r|} = \frac{1}{\Delta_f}H\left(\frac{\Delta_f}{2} - |r|\right) \quad (6)$$

This study considers various SGS models, including the Constant Smagorinsky (CS) [46], Dynamic Smagorinsky (DS) Dynamic Wong-Lilly (DWL) 1.5-order Turbulent Kinetic Energy Deardorff (TKED) Equilibrium Heinz (EH) and Dynamic Heinz (DH) models [46-53]. Using each of these models, the residual (or SGS) stress, $\tau^R$, can be determined as described in Table 1.

| Tag | Model's Name | Residual (or SGS) Stress Equation | |
|---|---|---|---|
| C1 | CS [46] | $\tau^{R,CS} = -2(C_{CS}\Delta_f)^2 \left|\frac{\partial \bar{u}}{\partial x}\right|\left(\frac{\partial \bar{u}}{\partial x}\right)$ | (7) |
| C2 | DS [46] | $\tau^{R,DS} = -2C_{DS}\Delta_f^2 \left|\frac{\partial \bar{u}}{\partial x}\right|\left(\frac{\partial \bar{u}}{\partial x}\right)$ | (8) |
| C3 | DWL [47, 48] | $\tau^{R,DWL} = -2C_{DWL}\Delta_f^{4/3}\left(\frac{\partial \bar{u}}{\partial x}\right)$ | (9) |
| C4 | TKED [49-51] | $\tau^{R,TKED} = -2K_{TKED}\left|\frac{\partial \bar{u}}{\partial x}\right|\left(\frac{\partial \bar{u}}{\partial x}\right)$ | (10) |
| C5 | EH [52, 53] | $\tau^{R,EH} = -2C_{EH}\Delta_f^2 \left|\frac{\partial \bar{u}}{\partial x}\right|\left(\frac{\partial \bar{u}}{\partial x}\right)$ | (11) |
| C6 | DH [52, 53] | $\tau^{R,DH} = -2K_{DH}\left(\frac{\partial \bar{u}}{\partial x}\right)$ | (12) |

**Table 1: Different residual (or SGS) models and their equations.**

In equation (7), the value of $C_{CS}$ typically ranges from 0.1 to 0.2. Here, this coefficient is set to 0.17 as Pope suggested based on the results obtained by Lilly [43,48]. The CS model is straightforward and easy to implement, offering a reasonable approximation for many flows. However, it can be overly dissipative, particularly in regions with low turbulence. Additionally, assuming a constant coefficient like $C_{CS}$ may not be appropriate for all flow conditions. To address certain limitations, the DS model has been modified. Its strengths include adaptive eddy viscosity, reduced numerical dissipation, improved accuracy in complex flows, and robustness across various conditions. However, its implementation can be complex, it is sensitive to filter choice, and may require calibration for specific flow conditions. Additionally, the model may struggle in very low Reynolds number flows and carries the risk of overfitting, which could lead to unphysical behavior in certain regions [44]. In equation (8), the DS coefficient, $C_{DS}$, is locally adjusted based on the resolved scales of turbulence. Its formulation can be represented as follows [46]:

$$C_{DS}^2 = \frac{\langle L_{11}M_{11}\rangle}{\langle M_{11}M_{11}\rangle} \quad (13)$$

Where $L_{11}$ and $M_{11}$ refer to the resolved stress (also known as the Germano identity) and the scaled composite rate-of-strain, respectively [54]. These quantities can be defined as follows:

$$L_{11} = \widetilde{\bar{u}^2} - \widetilde{\bar{u}}\widetilde{\bar{u}} \quad (14)$$

$$M_{11} = -2\Delta_f^2 \left(4\left|\frac{\widetilde{\partial \bar{u}}}{\partial x}\right|\left(\frac{\widetilde{\partial \bar{u}}}{\partial x}\right) - \left|\widetilde{\frac{\partial \bar{u}}{\partial x}}\right|\left(\widetilde{\frac{\partial \bar{u}}{\partial x}}\right)\right) \quad (15)$$

In equations (14) and (15), the tilde symbol indicates that test filtering is performed at a scale of $\Delta_t = 2\Delta_f$. Thus, $L_{11}$ represents the SGS stresses occurring between $\Delta_f$ and $\Delta_t$, as described by Basu [41].

The DWL model excels in simulating turbulent flows due to its ability to adaptively adjust model coefficient, $C_{DWL}$ in equation (9), based on local flow conditions, enhancing accuracy. It effectively captures a wide range of turbulence scales and handles complex geometries well. However, its computational demands can be high, making simulations resource-intensive. Additionally, it may face challenges in highly anisotropic or non-equilibrium flows, potentially leading to inaccuracies under such conditions [44]. The coefficient of DWL model, $C_{DWL}$, can be obtained by [47]:

$$C_{DWL} = \frac{\langle L_{11}N_{11}\rangle}{\langle N_{11}N_{11}\rangle} \quad (16)$$

Here, $N_{11}$ signifies the scaled gradient of velocity and can be expressed as follows (Basu [40]):

$$N_{11} = -2\Delta_f^{\frac{4}{3}}\left|\frac{\widetilde{\partial \bar{u}}}{\partial x}\right|\left(2^{\frac{4}{3}} - 1\right) \quad (17)$$

In both equations (13) and (16) the angular brackets indicate averaging across the entire one-dimensional domain. To prevent numerical instabilities, the dynamic coefficients of $C_{DS}^2$ and $C_{DWL}$ are assigned a value of zero whenever equations (13) and (16) produce a negative result. This widely adopted practice is referred to as "clipping" [40,45].

The 1.5 TKED model is efficient and straightforward, effectively simulating buoyancy-driven turbulence in atmospheric and oceanic flows. Its strengths include robust performance across various applications. However, it simplifies turbulence physics, may struggle with rapidly changing conditions, and can be sensitive to parameter choices, which may require careful calibration for accurate results [44]. The coefficient of this model in equation (10), denoted as $K_{TKED}$, can be computed as follows [51]:

$$K_{TKED} = 0.1\Delta_f k^{1/2} \quad (18)$$

According to equation (18), the value of SGS turbulent kinetic energy, $k$, is needed. To determine this, the following equation is used [51]:

$$\frac{\partial k}{\partial t} = -\frac{\partial(\bar{u}k)}{\partial x} + 2K_{TKED}\left[\left|\frac{\partial\bar{u}}{\partial x}\right|\left(\frac{\partial\bar{u}}{\partial x}\right)\right]^2 + 2\frac{\partial}{\partial x}\left[K_{TKED}\frac{\partial k}{\partial x}\right] - 0.7\frac{k^{3/2}}{\Delta_f} \quad (19)$$

The Heinz SGS model in LES offers notable strengths. Its key strength lies in its improved stability due to realizability constraints, which ensure the physical plausibility of the SGS stress tensor, reducing numerical instabilities. The dynamic adjustment of model coefficients based on local flow conditions enhances adaptability and accuracy without the need for empirical damping functions. However, the model has limitations. The increased computational complexity due to dynamic adjustments results in higher computational costs. Sensitivity to model parameters and initial conditions can affect robustness, necessitating careful calibration. Implementation can be challenging due to its unique approach, and further validation is needed across a broader range of turbulent flows to fully establish its reliability and generalizability [44]. The EH model, as described in equation (11), features a dynamic coefficient, $C_{EH}$. This coefficient can be determined by [53]:

$$C_{EH} = -\frac{L_{11}M_{11}^*}{M_{11}^*M_{11}^*} \quad (20)$$

Where $M^*_{11}$ is essentially $M_{11}$, excluding the first term, $4\left|\widetilde{\frac{\partial\bar{u}}{\partial x}}\right|\left(\widetilde{\frac{\partial\bar{u}}{\partial x}}\right)$, and replacing $\Delta_f$ with $\Delta_t$.

The coefficient of DH model in equation (12), denoted as $K_{DH}$, can be computed as follows [53]:

$$K_{DH} = C_{DH}\Delta_f k^{1/2} \quad (21)$$

By comparing relations (18) and (21), it is evident that in DH model, the coefficient of 0.1 is no longer present. Instead, the coefficient of $C_{DH}$ is dynamically calculated using equation (22). Thus, this approach (DH model) is effectively a generalization of TKED model.

$$C_{DH} = -\frac{L_{11}O_{11}}{O_{11}O_{11}} \quad (22)$$

Here, $O_{11}$ can be expressed as follows:

$$O_{11} = \sqrt{2}\Delta_t L_{11}^{1/2}\left(\widetilde{\frac{\partial\bar{u}}{\partial x}}\right) \quad (23)$$

According to equation (21), the value of SGS turbulent kinetic energy, $k$, is needed. To determine this, the following equation is used [53]:

$$\frac{\partial k}{\partial t} = -\frac{\partial(\bar{u}k)}{\partial x} + \sqrt{2}K_{DH}\left|\frac{\partial\bar{u}}{\partial x}\right| + \frac{\partial}{\partial x}\left[(v + K_{DH})\frac{\partial k}{\partial x}\right] - \frac{k^{3/2}}{\Delta_f} \quad (24)$$

### 2.3 Numerical Solution
The schematic of computational grid is shown in Figure 1. All quantities, including filtered velocity and SGS turbulent kinetic energy, are stored at the cell centers. According to Amani and Morinishi et al. two fully conservative, higher-order Finite Difference Methods (FEMs) can be applied to the spatial terms outlined in sections 2.1 and 2.2 [44,55]. These discretization schemes, as summarized in Table 2, can be used for a general variable such as $\phi$, which could be replaced with quantities like filtered velocity, $\bar{u}$, or SGS turbulent kinetic energy, $k$. Additionally, Table 3 presents three different discretization methods for the time derivative term [56-59].

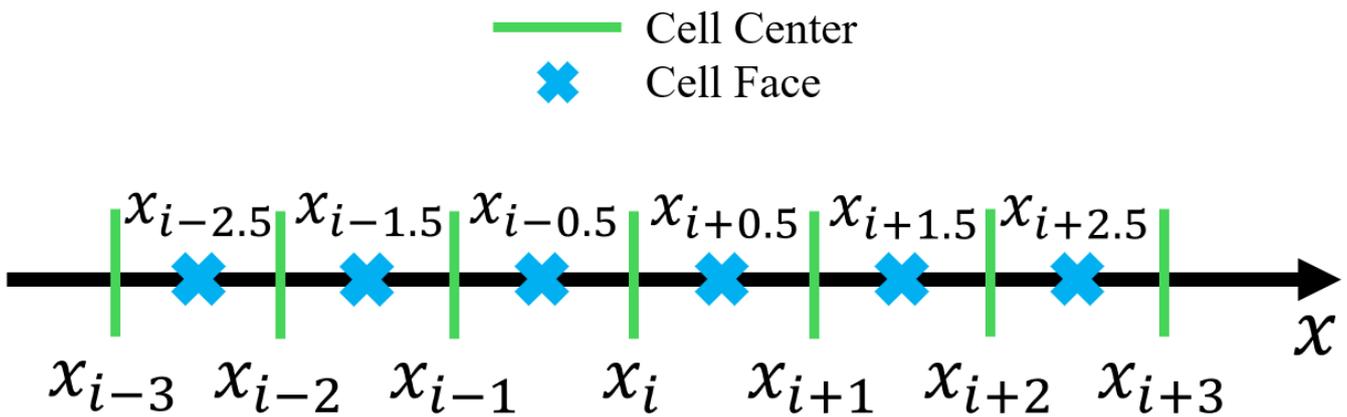

**Figure 1:** The Computational Grid, Cell Centers and Faces Highlighted with Green Line and Blue Symbol, Respectively.

| Term | Second-Order Discretization Schemes | |
|---|---|---|
| | Equation | |
| Gradient | $\frac{\partial \phi}{\partial x} = \frac{\phi_{i+1} - \phi_{i-1}}{2\Delta x}$ | (25) |
| Convection | $\phi \frac{\partial \phi}{\partial x} = \frac{1}{2} \frac{\partial \phi \phi}{\partial x} = \frac{1}{2}\left[\frac{(\phi_{i+1}+\phi_i)^2 - (\phi_i+\phi_{i-1})^2}{4\Delta x}\right]$ | (26) |
| Diffusion | $\frac{\partial^2 \phi}{\partial x^2} = \frac{\phi_{i+1} - 2\phi_i + \phi_{i-1}}{(\Delta x)^2}$ | (27) |
| **Term** | **Fourth-Order Discretization Schemes** | |
| | Equation | |
| Gradient | $\frac{\partial \phi}{\partial x} = \frac{4}{3}\left(\frac{\phi_{i+1}-\phi_{i-1}}{2\Delta x}\right) - \frac{1}{3}\left(\frac{\phi_{i+2}-\phi_{i-2}}{4\Delta x}\right)$ | (28) |
| Convection | $\phi \frac{\partial \phi}{\partial x} = \frac{1}{2}\frac{\partial \phi \phi}{\partial x} = \frac{1}{2}\left[\frac{4}{3}\left(\frac{(\phi_{i+1}+\phi_i)^2-(\phi_i+\phi_{i-1})^2}{4\Delta x}\right) - \frac{1}{3}\left(\frac{(\phi_{i+2}+\phi_i)^2-(\phi_i+\phi_{i-2})^2}{8\Delta x}\right)\right]$ | (29) |
| Diffusion | $\frac{\partial^2 \phi}{\partial x^2} = \frac{4}{3}\left(\frac{\phi_{i+1}-2\phi_i+\phi_{i-1}}{(\Delta x)^2}\right) - \frac{1}{3}\left(\frac{\phi_{i+2}-2\phi_i+\phi_{i-2}}{(2\Delta x)^2}\right)$ | (30) |

**Table 2:** Two Fully Conservative, Higher-Order FEMs for Spatial Terms [44,55].

| Order (Name) | Three Different Discretization Schemes | |
|---|---|---|
| | Equation | |
| First-Order (Explicit Euler) [56] | $\frac{\partial \phi}{\partial t} = \text{RHS} \rightarrow \phi^{n+1} = \phi^n + \Delta t (\text{RHS})^n$ | (31) |
| Second-Order (Adams-Bashforth) [57] | $\frac{\partial \phi}{\partial t} = \text{RHS} \rightarrow \phi^{n+2} = \phi^{n+1} + \Delta t \left[\frac{3}{2}(\text{RHS})^{n+1} - \frac{1}{2}(\text{RHS})^n\right]$ | (32) |
| Fourth-Order (Runge-Kutta) [58] | $\frac{\partial \phi}{\partial t} = \text{RHS} \rightarrow \phi^{n+1} = \phi^n + \frac{1}{6}\Delta t[C_1 + 2C_2 + 2C_3 + C_4]$ where: $\begin{cases} C_1 = (\text{RHS})^n \\ C_2 = \left(\text{RHS} + \frac{C_1}{2}\right)^{n+\frac{\Delta t}{2}} \\ C_3 = \left(\text{RHS} + \frac{C_2}{2}\right)^{n+\frac{\Delta t}{2}} \\ C_4 = (\text{RHS} + C_3)^{n+\Delta t} \end{cases}$ | (33) |

**Table 3:** Three Different Discretization Methods for the Time Derivative Term.

## 3. Validation

Before analyzing the LES solution, it is essential to validate the accuracy of the Python code and discretization schemes. To achieve this, two validation cases, as referenced in Moin are first considered [60]. The validation cases include the Linear Advection-Diffusion (LAD) equation and the Non-Linear Burgers (NLB) equation, both of which possess analytical solutions under laminar flow conditions. For both cases, Table 4 presents the governing equations along with the corresponding boundary and initial conditions.

| Name | Validation Cases Under Laminar Flow Conditions | |
|---|---|---|
| | Equation, Boundary Condition (B.C.), and Initial Condition (I.C.) | |
| Linear Advection-Diffusion (LAD) [60] | $\frac{\partial u}{\partial t} + \frac{\partial u}{\partial x} = 0.05 \frac{\partial^2 u}{\partial x^2}$ on $0\,m \leq x \leq 1\,m$ for $t = 0.75s$ <br> B.C.: $u(0,t) = u(1,t)$ <br> I.C.: $u(x,0) = \begin{cases} 1 - 25(x - 0.2)^2 & ; 0\,m \leq x < 0.4\,m \\ 0 & ; \text{otherwise} \end{cases}$ | (34) |
| Non-Linear Burgers (NLB) [60] | $\frac{\partial u}{\partial t} + u\frac{\partial u}{\partial x} = \frac{\partial^2 u}{\partial x^2}$ on $0\,m \leq x \leq 2\pi\,m$ for $t = 0.6s$ <br> B.C.: $u(0,t) = u(2\pi,t)$ <br> I.C.: $u(x,0) = 10\sin(x)$ | (35) |

**Table 4:** The validation cases: governing equations along with the boundary and initial conditions [60].

It is important to note that in both cases, the time step size, $\Delta t$, is set to 0.005s. Additionally, the value of 16 is used for the number of grid points, $N$. The velocity profiles for both cases at various time points are compared with the analytical results and are presented in Figure 2a-d [60].

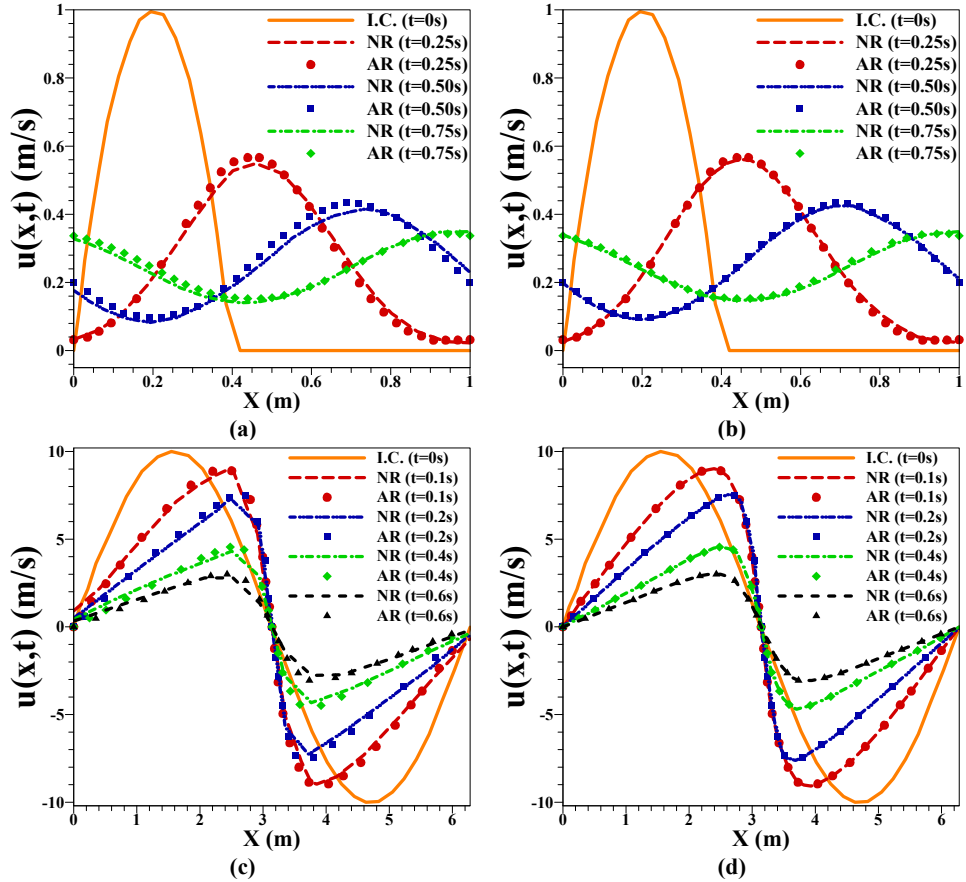

**Figure 2:** Velocity profiles at different time points versus *x*, (a) LAD using a second-order scheme, (b) LAD using a fourth-order scheme, (c) NLB using a second-order scheme, and (d) NLB using a fourth-order scheme. The legend includes AR NR (Numerical Results), and I.C. (Initial Condition) [60].

According to Figure 2a-d, the numerical results from the fourth-order scheme align perfectly with the analytical solutions. However, in both validation cases, the second-order scheme shows slight deviations from the analytical results. When comparing spatial fourth-order discretization scheme to second-order discretization scheme in the context of solving PDEs like the LAD equation and the NLB equation, several factors come into play:

Firstly, in numerical discretization, the accuracy and convergence rate of a scheme are closely tied to its order. For second-order discretization, the truncation error is $O(\Delta x^2)$. This means that when the grid spacing, $\Delta x$, is halved, the error decreases by a factor of 4, but the error remains relatively large compared to higher-order schemes. In contrast, fourth-order discretization has a truncation error of $O(\Delta x^4)$. Halving the grid spacing in this case reduces the error by a factor of 16, resulting in a significantly more accurate solution for the same grid size (see Table 2). Secondly, LAD equation often models the transport of a quantity like heat or pollutants. For advection-dominated problems (where the advection term is much stronger than the diffusion term), numerical schemes with lower-order discretization can introduce artificial dispersion (phase errors) and dissipation (amplitude damping) as shown in Figure 2a. A fourth-order scheme reduces both these effects significantly compared to a second-order scheme, leading to more accurate wave propagation and sharper gradients in the solution (refer to Figure 2b). Furthermore, NLB equation is used to model shock waves and turbulence, where nonlinear effects dominate. Higher-order discretization can better capture the steep gradients and complex structures typical of such solutions. Fourth-order methods reduce numerical dissipation, preserving sharper discontinuities (shocks) and yielding a more accurate representation of the solution (refer to Figure 2c-d). Thirdly, second-order discretization generally necessitates a finer grid to achieve the same level of accuracy as a fourth-order scheme (refer to Figure 2a-d). This finer grid increases the number of grid points, thereby raising the computational cost. Although fourth-order schemes offer greater accuracy, they involve a wider stencil, incorporating more grid points in the derivative approximation (see Table 2). While this broader stencil can slightly elevate computational complexity, the enhanced accuracy typically justifies the additional expense.

## 4. Results and Discussion
### 4.1 Dynamic Coefficients
This section investigates the effects of higher-order spatial discretization schemes on the dynamic coefficients of various SGS models within the framework of LES. Figure 3a-d illustrates the time evolution of the dynamic coefficients for the DS, DWL, EH, and DH models. Notably, as indicated in Figure 3a-b, the dynamic SGS coefficients for the DS and DWL models exhibit qualitatively similar patterns over time, as described by Basu [40]. The most significant variations in the coefficients occur during five specific intervals: 35-40 s, 65-70 s, 135-140 s, 160-165 s, and 175-180 s. The second-order discretization scheme amplifies these changes considerably more than the fourth-order scheme. For example, the maximum values of the coefficients within the specified intervals, as detailed in Table 5, underscore the differences between the two numerical methods. The second-order scheme typically introduces more numerical dissipation than the fourth-order scheme, which tends to dampen high-frequency components (small scales) in the solution. In the context of LES, this additional dissipation can influence the resolved scales, leading to a greater dependence on the SGS model to represent energy transfer between the resolved and subgrid scales. Consequently, the dynamic coefficients may increase to compensate for this effect, resulting in the observed rise. Moreover, the dynamic models adapt their coefficients based on local flow characteristics and the interaction between resolved and subgrid scales. With a higher-order scheme, the resolved scales are represented more accurately, potentially leading to a smaller SGS coefficient. It is noteworthy that the results for the DS and DWL coefficients using the fourth-order scheme align perfectly with those obtained by the spectral method [41].

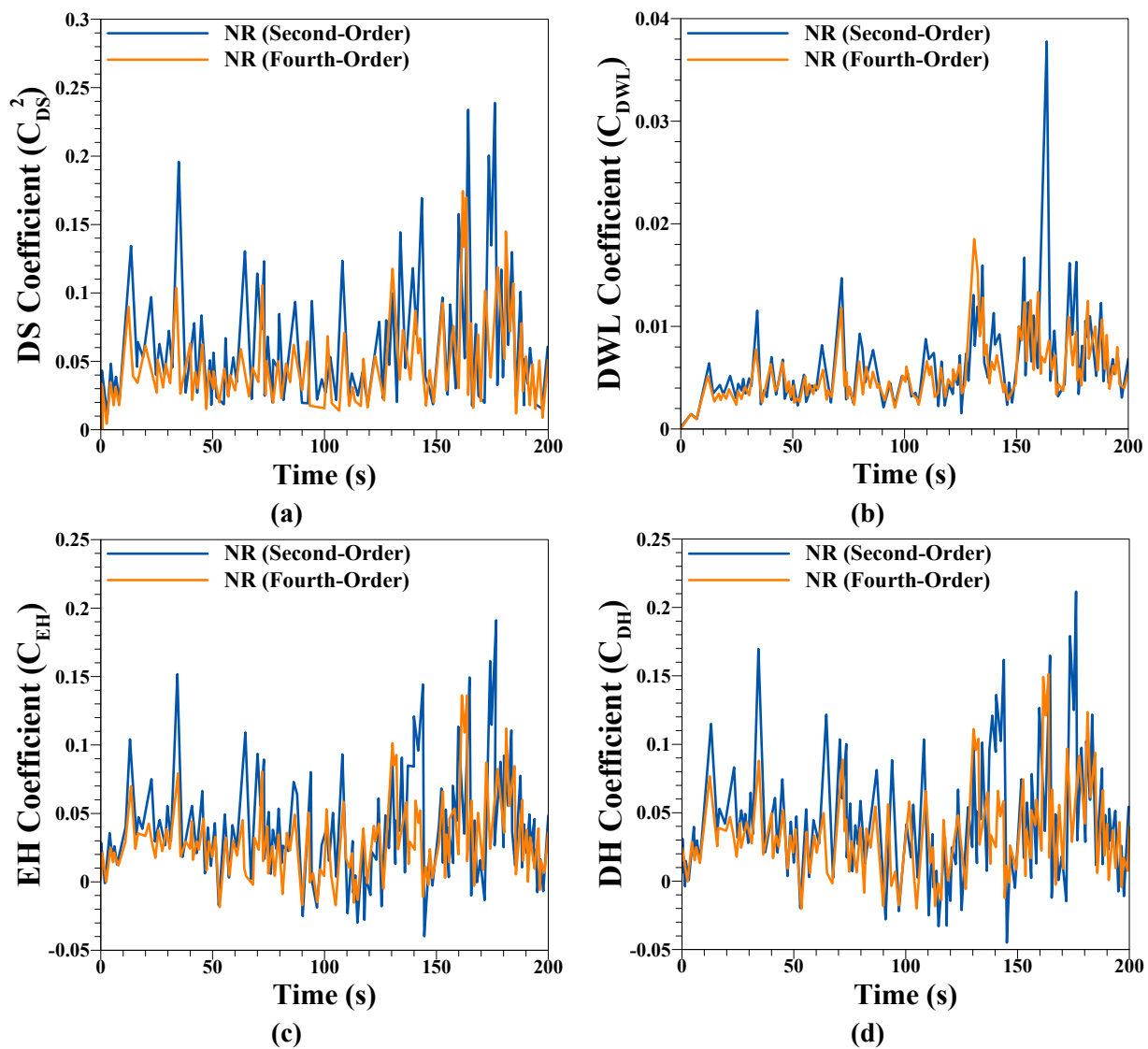

**Figure 3:** Time evolution of dynamic coefficient distributions for various SGS models: (a) DS, (b) DWL, (c) EH, and (d) DH. The legend includes Numerical Results (NR) for both numerical schemes.

| Time Interval (s) | DS Model (Order) | | DWL Model (Order) | | EH Model (Order) | | DH Model (Order) | |
|---|---|---|---|---|---|---|---|---|
| | (2nd) | (4th) | (2nd) | (4th) | (2nd) | (4th) | (2nd) | (4th) |
| 35-40 | 0.196 | 0.104 | 0.012 | 0.008 | 0.152 | 0.079 | 0.170 | 0.088 |
| 65-70 | 0.131 | 0.059 | 0.015 | 0.012 | 0.109 | 0.045 | 0.122 | 0.049 |
| 135-140 | 0.118 | 0.087 | 0.011 | 0.009 | 0.144 | 0.059 | 0.162 | 0.065 |
| 160-165 | 0.234 | 0.169 | 0.038 | 0.006 | 0.149 | 0.136 | 0.165 | 0.149 |
| 175-180 | 0.239 | 0.119 | 0.016 | 0.006 | 0.191 | 0.082 | 0.179 | 0.097 |

**Table 5: The comparison of two spatial discretization schemes across four SGS model coefficients over five time intervals.**

In LES, turbulence is divided into large-scale motions that are resolved explicitly and small-scale motions that are modeled. SGS models typically assume forward scatter, where energy transfers from large to small scales. However, under specific conditions, such as highly anisotropic flows or organized structures, backscatter can occur. When SGS coefficients turn negative, as seen in Figure 3c-d, it indicates backscatter, where small-scale energy feeds back into large scales. This phenomenon often manifests in regions with organized turbulence like shear layers or coherent structures [44]. Comparing Figure 3c-d with Figure 3a-b reveals that EH and DH models excel in accurately predicting energy exchange in scenarios involving backscatter. Moreover, the coefficients of the DS and DWL methods have never been negative (refer to Figure 3a-b), due to the clipping method employed in section 2.2 to enhance stability in the numerical solution. Consequently, these methods lack the capability to predict any backscatter in this study.

### 4.2 Distributions of Velocity

This section explores two key impacts on the distribution of turbulent flow velocity: the effects of various SGS modeling approaches and the influence of spatial discretization schemes. Figure 4a-f presents the velocity distributions as a function of x at the final time of t=200s for various SGS models. Figure 4a-f demonstrates that, despite the variety of SGS models used, they all produced similar velocity distributions in the context of the one-dimensional SBE simulation which is consistent with the findings of Basu [41]. This consistency can be attributed to several factors. In one-dimensional turbulence, the flow dynamics are significantly simplified compared to higher dimensions. This reduction in complexity can diminish the differences between SGS models, particularly when the main role of these models is to represent the effects of smaller scales in more complex, multi-dimensional turbulence. In one-dimensional, the role of SGS models might be less critical, leading to similar outcomes. Additionally, the Burgers equation includes a dissipative term (the viscosity term) that inherently influences the dynamics, particularly at longer timescales (for example, at a final time of t=200s). This term smooths out variations in the velocity field, as all the SGS models employed are designed to represent dissipation at smaller scales. Consequently, they may converge on similar dissipation characteristics over time, leading to comparable velocity distributions (refer to Figure 4a-f). Moreover, if the turbulence has reached a state where energy input and dissipation are balanced, the specific details of the SGS model may exert minimal influence on the large-scale velocity distribution. In such scenarios, the different SGS models may yield similar energy dissipation rates, resulting in nearly identical outcomes.

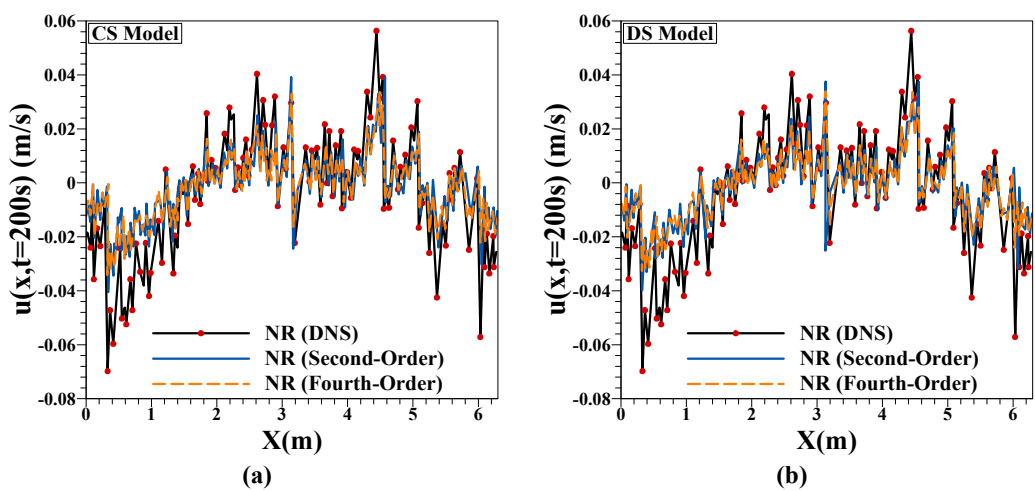

(a) (b)

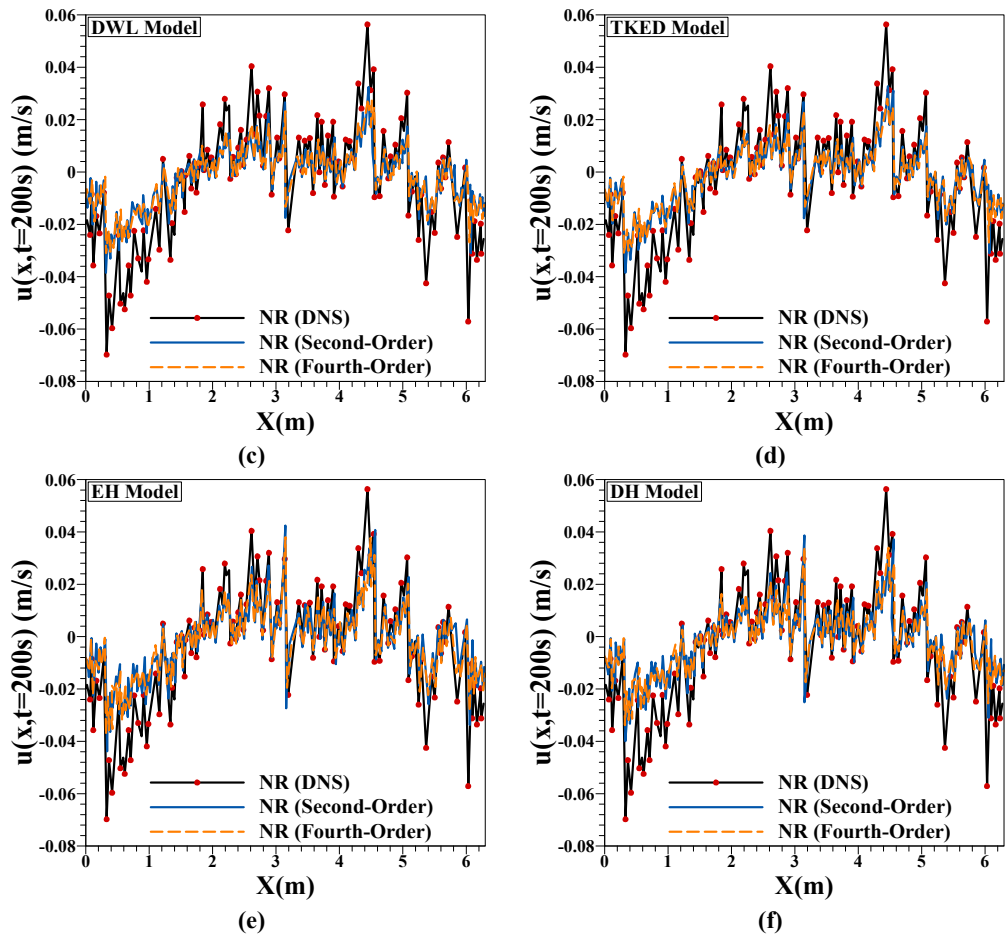

**Figure 4:** Velocity profiles versus x at the final time of t=200s for various SGS models: (a) CS, (b) DS, (c) DWL, (d) TKED, (e) EH, and (f) DH. The legend also includes Numerical Results (NR) representing different numerical schemes.

Both the CS and DS models utilize a similar eddy viscosity mechanism, with the dynamic version adapting based on the flow conditions. However, in a one-dimensional context, such adjustments might be less significant, as indicated in Figure 4a-b, consistent with the findings of Basu [41]. The DWL model, another eddy viscosity approach, incorporates modifications aimed at improving the handling of anisotropic turbulence. Yet, these enhancements may have limited impact in a one-dimensional setting, as illustrated in Figure 4c, as described by Basu [41]. The TKED model, which involves a prognostic equation for turbulent kinetic energy, may not differ significantly from other models in a one-dimensional scenario, as seen in Figure 4d. The added complexity of this model could be superfluous in such a simplified case. Lastly, the EH and DH models, grounded in the statistical theory of Probability Density Functions (PDFs), exhibit greater realizability. This is particularly evident in Figure 4e-f, where they more accurately capture the fluctuations in velocity distribution. In LES, the choice of SGS model can sometimes overshadow the effects of numerical discretization, especially in more complex flows. However, according to Figure 4a-f, where different SGS models already produced similar results, it suggests that the SGS effects might be well-represented even with the lower-order scheme, making the difference between the second-order and fourth-order schemes less noticeable. Therefore, in the following section, only the results from the fourth-order discretization scheme will be presented.

### 4.3 Distributions of Resolved-Scale Kinetic Energy

This section examines the impact of different SGS modeling approaches on resolved-scale kinetic energy within the framework of LES. Figure 5 depicts the time evolution of resolved-scale kinetic energy for several SGS models, including DS, DWL, and DH. Additionally, a specific scenario is considered in which equation (4) is solved by setting $\tau^R=0$, as reflected in the "No Model" results shown in Figure 5. The simulation without any SGS model exhibits significantly higher levels of random fluctuations, as is evident in Figure 5. The resolved-scale kinetic energy is markedly elevated in the "No Model" simulation. This increase results from the accumulation of energy due to the absence of SGS dissipation which is consistent with the findings of Basu [40]. This undesirable effect becomes more pronounced in LES simulations with coarser resolutions (not shown here). In section 4.2, it is observed that the velocity distribution shows similar results across different SGS models. However, when comparing the resolved-scale kinetic energy to DNS results in Figure 5, the DS and DWL models perform better than the DH model. The differences between SGS models in this regard arise from several factors. First, DS model adjusts the Smagorinsky coefficient dynamically based on the local flow conditions. It tends to be more adaptable and can better match the dissipation rates in regions of high and low turbulence, leading to a more accurate energy cascade and better alignment with DNS results. Similar to the DS model, this model dynamically adjusts parameters but with different assumptions and modifications in the formulation as stated in section 2.2, which might lead to different responses to the local turbulence structure. The DH model focuses on reproducing the correct dissipation and backscatter behavior (refer to Figure 3d), which might not align as well with the specific energy distribution observed in DNS. According to Figure 5, it might over- or under-estimate certain contributions to the kinetic energy, leading to a deviation from DNS results. Second, the DS and DWL models might be more sensitive to small-scale structures in the flow, which are crucial for accurately capturing the energy dissipation rates and ensuring that the energy at the resolved scales matches DNS results, as described by Basu [40]. On the other hand, the DH model might not capture these small-scale structures as effectively, leading to a less accurate prediction of the resolved-scale kinetic energy, as depicted in Figure 5.

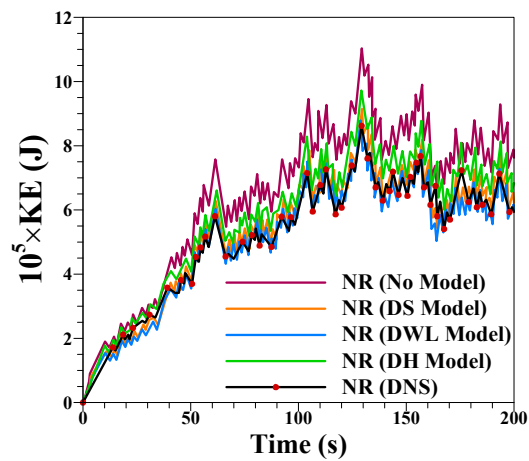

**Figure 5:** Time evolution of resolved-scale kinetic energy distributions for various SGS models.

The legend includes Numerical Results (NR) for numerical modeling.

## 5. Conclusion

In this study, fully conservative higher-order schemes were implemented for LES of stochastic Burgers turbulence, rigorously validated against analytical solutions for the linear advection-diffusion equation and the nonlinear Burgers equation under laminar conditions. These validated schemes were then used to investigate the impact of different spatial discretization schemes on the dynamic coefficients of SGS models. The findings reveal that the choice of spatial discretization significantly affects the behavior of dynamic coefficients in various SGS models within LES. The study highlights that second-order discretization amplifies fluctuations in dynamic SGS coefficients more than fourth-order discretization, particularly during key time intervals. This amplification is attributed to the higher numerical dissipation introduced by the second-order scheme, which affects the resolved scales and requires greater contribution from the SGS model to represent energy transfer between scales. Conversely, the fourth-order scheme, with its ability to more accurately capture resolved scales, results in smaller dynamic coefficients. The DS and DWL models exhibit similar temporal patterns in their coefficients, indicating the robustness of these methods when employing a higher-order scheme. The study also explores the effects of various SGS models and spatial discretization schemes on the turbulent flow velocity distribution in a one-dimensional SBE simulation. Despite using different SGS models—CS, DS, DWL, TKED, EH, and DH—the results show similar velocity distributions at the final time. This consistency is likely due to the simplified dynamics of one-dimensional turbulence, where differences between SGS models diminish, especially since their primary function is to model the effects of smaller scales in more complex, multi-dimensional turbulence. The inherent dissipative term in the Burgers equation further smooths out variations in the velocity field, leading to convergence in dissipation characteristics across different SGS models over time.

The analysis suggests that in a one-dimensional turbulence scenario, the choice of SGS model has minimal influence on the large-scale velocity distribution, as energy input and dissipation reach a balance. This is evident in the similar outcomes produced by various SGS models, which rely on different eddy viscosity mechanisms and statistical theories. While numerical discretization can significantly influence LES outcomes in more complex flows, in this one-dimensional case, the effects of SGS modeling may be well-represented even with a lower-order scheme, making the difference between second-order and fourth-order schemes less pronounced. Additionally, the study reveals notable differences in how various SGS models impact resolved-scale kinetic energy. Simulations without any SGS model show a significant increase in resolved-scale kinetic energy due to the lack of SGS dissipation, which allows unchecked energy accumulation, especially at coarser resolutions. The DS and DWL models, which adapt their parameters based on local flow conditions, align more closely with DNS results compared to the DH model. The adaptability of the DS and DWL models enables them to match dissipation rates more accurately across different turbulence intensities, ensuring a more precise energy cascade. In contrast, the DH model, despite its focus on reproducing correct dissipation and backscatter behavior, may struggle to capture small-scale structures effectively, leading to discrepancies in resolved-scale kinetic energy compared to DNS results. These differences underscore the importance of capturing small-scale turbulence structures in LES for accurate energy dissipation predictions, with the DS and DWL models demonstrating superior performance in this regard.